\documentclass[]{JFM-FLM_Au}
\lefttitle{S. Prakhar, J.-H. Seo and R. Mittal}
\righttitle{Journal of Fluid Mechanics}

\title{
From vortices to forces -- a data-driven framework for unsteady lift generation in three-dimensional vortex-dominated flows  
}
\author{Suryansh Prakhar\aff{1}, Jung-Hee Seo\aff{1} \and Rajat Mittal\aff{1}}

\affiliation{\aff{1}Department of Mechanical Engineering, Johns Hopkins University, Baltimore, MD, USA}

\corresau{Rajat Mittal, \email{mittal@jhu.edu}}

\begin{document}
\maketitle

\begin{abstract}
Time-varying flow-induced forces on bodies immersed in fluid flows play a key role across a range of natural and engineered systems, from biological locomotion to propulsion and energy-harvesting devices. These transient forces often arise from complex, dynamic vortex interactions and can either enhance or degrade system performance. However, establishing a clear causal link between vortex structures and force transients remains challenging, especially in high-Reynolds number nominally three-dimensional flows. In this study, we investigate the unsteady lift generation on a rotor blade that is impulsively started with a span-based Reynolds number of 25,500. The lift history from this direct-numerical simulation reveals distinct early-time extrema associated with rapidly evolving flow structures, including the formation, evolution, and breakdown of leading-edge and tip vortices. To quantify the influence of these vortical structures on the lift transients, we apply the force partitioning method (FPM) that quantifies the surface pressure forces induced by vortex-associated effects. Two metrics -- $Q$-strength and vortex proximity -- are derived from FPM to provide a quantitative assessment of the influence of vortices on the lift force. This analysis confirms and extends qualitative insights from prior studies, and offers a simple-to-apply data-enabled framework for attributing unsteady forces to specific flow features, with potential applications in the design and control of systems where unsteady aerodynamic forces play a central role.
\end{abstract}


\section{Introduction}
The generation of time-varying flow-induced forces on bodies immersed in flows is central to a multitude of flow problems ranging from biological and bioinspired locomotion in fluids~\citep{birch2004force,lentink2009rotational,seo2022improved} to lifting surfaces such as wings \citep{lambert2019leading,jardin2021empirical} and rotors \citep{nabawy2017role,raghav2015advance}, and thrust (propulsors, propellers, etc.) and power (pumps, turbines, etc.) generating devices. The time-variation in these forces can result from intrinsic variations in the flow (such as through vortex shedding, turbulence etc), from extrinsic effects (such as acceleration/deceleration of the body) or through the  combined effect of both. Understanding the mechanisms and/or flow features that give rise to these transient forces is of critical importance since the transients can either be desirable (i.e. transient lift forces are key to high performance of flapping wings/fins~\citep{birch2004force,lentink2009rotational,seo2022improved}) or undesirable (i.e. transient forces can generate flow noise~\citep{brentner2003modeling,mfpm_2025}) and drive flow-induced vibrations/flutter~\citep{williamson2004vortex}. Methods than can enable an understanding of the causality of these force transients through analysis of simulation or experimental data could be used to inform design changes, operational modification or flow control strategies that accentuate (in the former case) or mitigate (in the latter) these effects.

\begin{figure}
\centering
\includegraphics[width=0.9\textwidth]{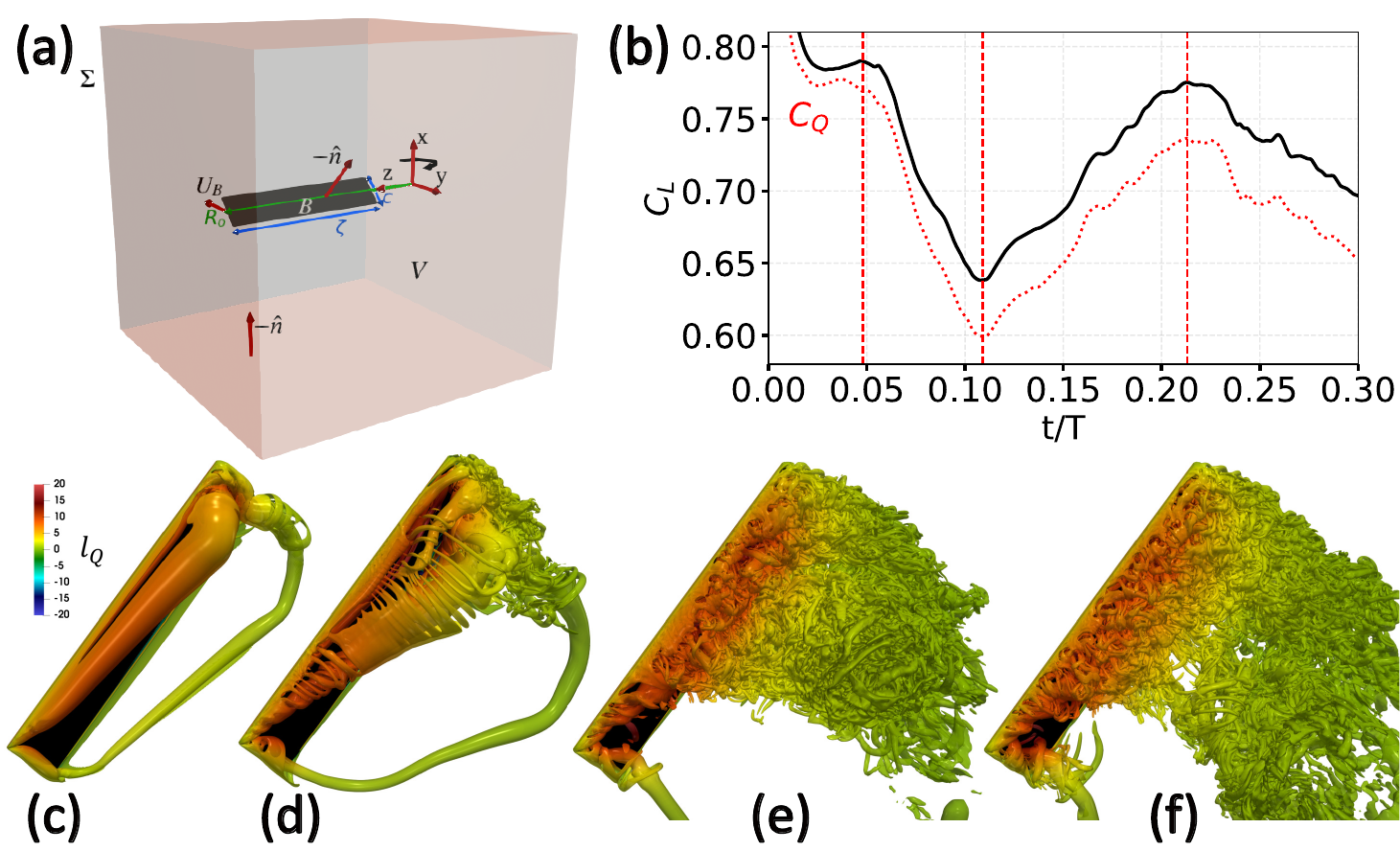}
\caption{(a) Flow schematic (not to scale) for the revolving blade showing the problem configuration with the origin shown at the center of revolution. (b) The pressure lift coefficient normalized based on the tip velocity and the blade area ($\frac{1}{2}\rho v_t^2 A_B$) is shown with the vertical red lines shown at $t/T=$ 0.048, 0.109, and 0.213 and the corresponding flow field shown in (c), (d) and (e) respectively. (f) Flow field shown at a later time, $t/T=0.29$. $T=2\pi/\Omega_z$ is the revolution period,  $C_Q=L_Q/(\frac{1}{2}\rho v_t^2 A_B)$ is the vortex-induced lift coefficient, and $L_Q$ is defined later in Eq. \ref{fpmEqn}.}
\label{fig:early_liftcoeff}
\end{figure}

Determining the source of these force transients is, however, challenging, particularly in three-dimensional configurations at moderate to high Reynolds number flows. These flows contain a wide range of highly dynamic vortex structures which can all influence the pressure loading to different degrees. Consider the flow of interest here: the flow associated with a impulsively started rotor blade at a span-based Reynolds number of 25,500 (figure \ref{fig:early_liftcoeff}(a)). Figure \ref{fig:early_liftcoeff}(b) shows the lift force versus time for about the first quarter revolution, and we note three distinct early-time extrema in the lift force; an initial maximum at $t/T$ = 0.048, a minima at $t/T$=0.109 and a second maximum at $t/T$=0.213. 

The vortex structures corresponding to these time-instances and $t/T$=0.29 are shown in figures \ref{fig:early_liftcoeff}(c--f). At $t/T$ = 0.048, we note a very coherent, initial leading-edge vortex (LEV) that increases in strength towards the blade tip. We also observe a small tip-vortex that connects to the starting vortex that is released from the trailing-edge into the wake. At $t/T$=0.109, the initial LEV is starting to separate from the outer regions of the blade and braid-vortices and other instabilities begin to manifest. A smaller LEV is also observed to form at this time. The situation at $t/T$=0.213 is difficult to parse since the flow has already transitioned to turbulence at this point and all the vortex structures are seen to merge in one conglomeration that extends from about 24\% span to the tip and into the wake. At the later time of $t/T$=0.29, the vortex structures look similar to those at $t/T$=0.213, but the lift has reduced by about 13\%. 

While we generally understand that the formation of a LEV enhances lift~\citep{polhamus1966concept,LEV_arev} and the separation of a LEV (also sometimes referred to as dynamic stall) diminishes lift \citep{tsang2008dynamic, LEV_arev}, we still lack \emph{quantitative} ways of verifying these concepts from data coming from simulations or experiments of complex 3D flows configurations.  One relevant study in this regard is that of \cite{jardin2021empirical}  who showed that reasonable estimates of the lift for accelerating wings could be obtained using the circulation and position of the LEV estimated from simulation data. However, it is not clear how their model, which they applied to a low Reynolds number (Re=500 airfoil) 2D flow, can be extended to much higher Reynold number flows over \emph{three-dimensional} bodies such as the rotor blade. \cite{otomo2021unsteady} presented a model that combined classic Theodorsen and thin-airfoil theories to predict the unsteady lift on a nominally two-dimensional flapping wing from wing kinematics only. However, they noted that the interaction of the late-forming TEV with the LEV generates errors in the model prediction. Furthermore, their vortex-lift correction requires the calculation of vortex circulation, which is not applicable to 3D configurations. Indeed in situations such as at $t/T$=0.213 and $t/T$=0.290, it is difficult to even clearly identify distinct vortex structures for which circulation can be estimated.

We therefore need a data-enabled framework that applies to complex flow fields emerging from high-Reynolds number, fully three-dimensional flows. Motivated by this, we employ the force partitioning method (FPM, see~\cite{menon2021a,menon2021b,menon2021c}) to develop a method to address precisely this issue -- the connection of the extrema in the lift of this impulsively accelerated blade to the vortical features of the flow. 

\section{Methodology}
\subsection{Flow Solver and Configuration}
We use our in-house flow solver called ViCar3D~\citep{mittal2008,seo2011} to solve the following set of incompressible Navier-Stokes equations written in a non-inertial rotating reference frame~\citep{speziale_rrf_eqn}:
\begin{equation}
     \nabla \cdot {\bf u} = 0 \,\,\, ,
     \label{conteqn}
 \end{equation}
 \begin{equation}
     \frac{\partial {\bf u}}{\partial t} + {\bf u}\cdot \nabla {\bf u}  = -\frac{1}{\rho}\nabla p +\nu \nabla^2 {\bf u} -2\boldsymbol{\Omega} \times {\bf u} - {\boldsymbol{\Omega} \times (\boldsymbol{\Omega} \times {\bf x})} \,\,\,,
\label{momeqn}
 \end{equation}
where, ${\bf u}$ is the flow velocity in the absolute frame, $\boldsymbol{\Omega}=[0,0,\Omega_z]$ is the reference frame rotation vector, $2\boldsymbol{\Omega} \times {\bf u}$ is the Coriolis acceleration term and ${\boldsymbol{\Omega} \times (\boldsymbol{\Omega} \times {\bf x})}$ is the centrifugal acceleration term. The solution is 2$^\text{nd}$-order accurate in both space and time \citep{mittal2008,seo2011}. A variety of cases where this solver has been used can be found in \cite{mittal_origin_2023,mittal_freeman}. 

The simulations employ a rotor blade which is a zero thickness, $AR=\zeta/c=5$ rectangular flat plate. 
The blade pitch angle is set to $45^\circ$ and the Reynolds number based on the span ($\zeta$) and tip-velocity ($v_t$) is 25,500.
The domain is $10R \times 10R \times 8R$, where $R$ is the outer radius of the blade, and the revolution center is placed at the middle of the domain. The blade initiates an impulsive revolution at $t=0$ with an angular velocity of $\Omega_z$ which remains constant afterwards.
Numerically, this results in an angular acceleration of $\Omega_z/\Delta t$ at $t=0$, where $\Delta t=0.0004/\Omega_z$ is the time step size used in the simulation.
A grid with 220 million points is used and the grid independence study is summarized in Appendix \ref{app:gc}. 

\subsection{Force Partitioning Method~(FPM)}
\label{sec:FPM}

\begin{figure}
\centering
\includegraphics[width=0.9\textwidth]{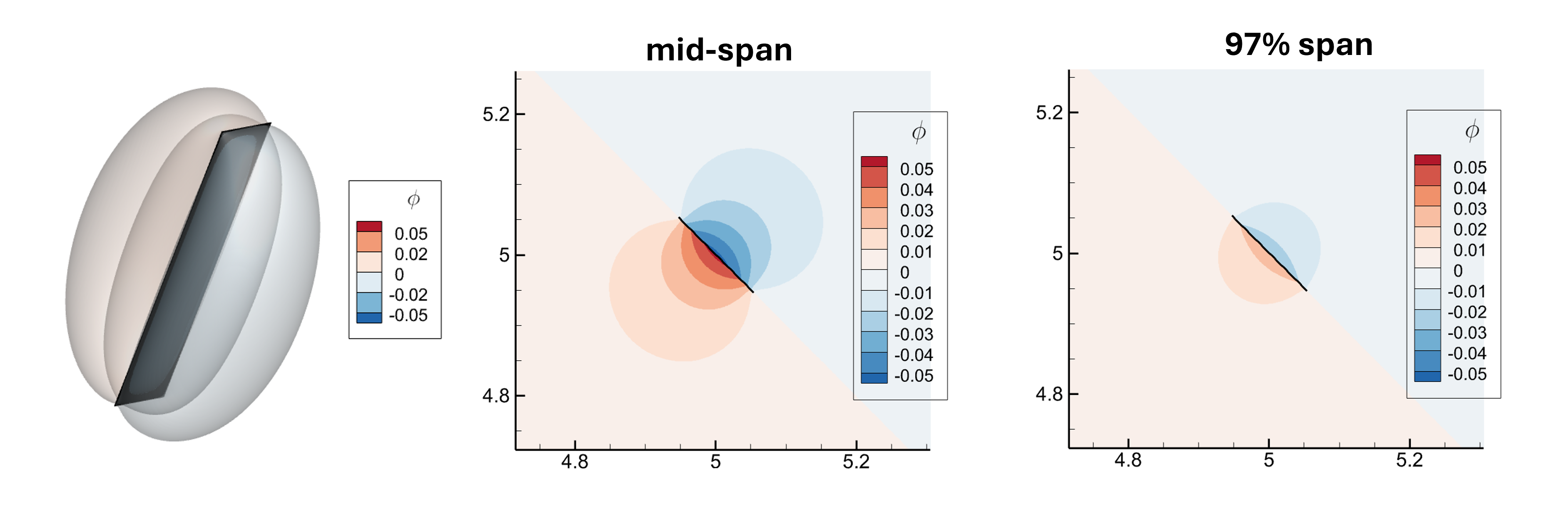}
\caption{The iso-surface of $\phi/R$ is shown on the left. Right - 2D contour of $\phi/R$ on two planes normal to the span - rotor mid span and near the tip (97\% span) are shown. }
\label{fig:phi_contour}
\end{figure}

The pressure forces dominate the forces on the rotor blade at these Reynolds number and the Force Partitioning Method (FPM;~\citet{zhang2015centripetal,menon2021c}) 
enables us to decompose the surface pressure force into components associated with vortices (vortex-induced force), the forces associated with acceleration reaction (a.k.a. the added mass forces) and pressure force due to viscous diffusion of momentum. The details of FPM and its implementation to immersed bodies can be found in~\cite{menon2021a,menon2021b,menon2021c} and application to this same rotor blade in~\cite{mfpm_2025}. 
In the current case, there are additional components due to the Coriolis and centrifugal force terms but these terms have no contribution on the pressure force if the domain is sufficiently large. 
We also note that the acceleration reaction can only contribute to the spanwise force, but is zero for the present rotor blade geometry (see Appendix \ref{app:gc}).
The numerical impulsive start with the angular acceleration of $\Omega_z/\Delta t$ at $t=0$ generates added mass lift at the first time-step, but this is zero for the subsequent time-steps. 
The viscous momentum diffusion component contributes a small but non-negligible (4.5\%) component to the lift, but overall, as in other high Reynolds number cases~\citep{seo2022improved,raut2024hydrodynamic} the vortex-induced force is by far the dominant contributor to lift generation (see Fig. \ref{fig:early_liftcoeff}(b)). Thus, FPM leads to the following expression for the pressure lift ($L$) on the rotor blade (see Fig. \ref{fig:early_liftcoeff}(a)) 
\begin{equation}
\begin{aligned}
   L =  \int_B p \, n_z dS \approx L_Q = -2\int_{V} \left(\rho Q \phi\right) dV = \int_{V} l_Q \, dV
    \label{fpmEqn}
\end{aligned}
\end{equation}
where $n_z$ is the vertical component of the surface unit normal pointing into the body and $B$ and $V$ are the body surface and fluid volume, respectively. $L_Q$ is the vortex-induced lift force, and $l_Q $ is referred to as the vortex-lift density and it is equal to $ - 2 \rho Q \phi$, where $Q = \frac{1}{2} (|| \boldsymbol{\omega}||^2- || {\bf S}||^2) \equiv -\frac{1}{2}{\bf \nabla} \cdot ({\bf u}\cdot {\bf \nabla u})$~\citep{jeong1995identification} and $\phi$ is the ``influence field'' in the direction of lift ($z$) which is obtained by solving the following equation
\begin{equation}
    \nabla^2\phi=0 \text { in } V;  \text{   with} \, \,     \nabla\phi \cdot {\bf n}=  \begin{cases}
    n_z & \text{on $B$},\\
    0 & \text{on $\Sigma$}\, ,
  \end{cases}
\end{equation}
where $\Sigma$ is the domain boundary. Figure \ref{fig:phi_contour} shows iso-surface of $\phi$ along with contours at two spanwise planes over the blade. Given Eq. \ref{fpmEqn}, we can estimate the pressure lift force induced on the surface from any volume segment in the fluid domain.

\section{Results}
\begin{figure}
\centering
\includegraphics[width=0.6\textwidth]{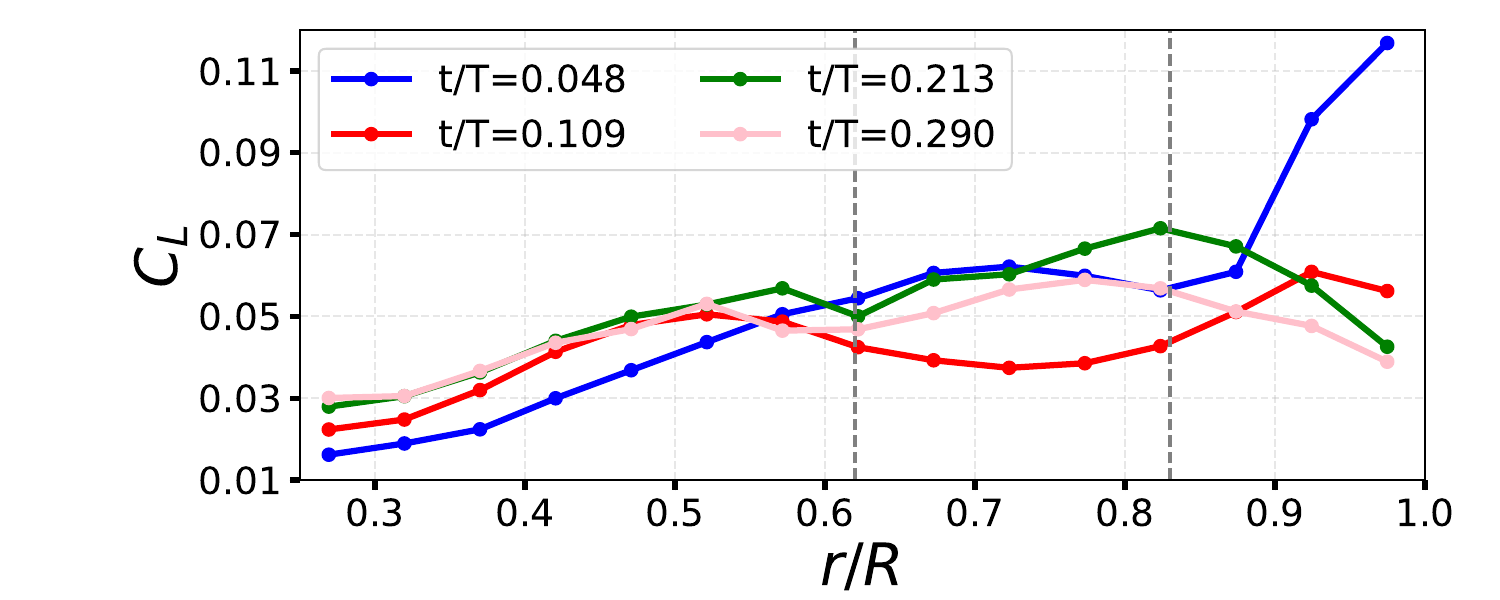}
\caption{Sectional lift coefficient at key time-instances. The rotor is divided into 15 parts along the span to show the contribution of each part to the lift force as the flow evolves. The vertical lines show the region over which spanwise average is computed for plots shown in figure \ref{fig:thin_slice_lift}.}
\label{fig:15slice_lift}
\end{figure}

\begin{figure}
\centering
\includegraphics[width=0.7\textwidth]{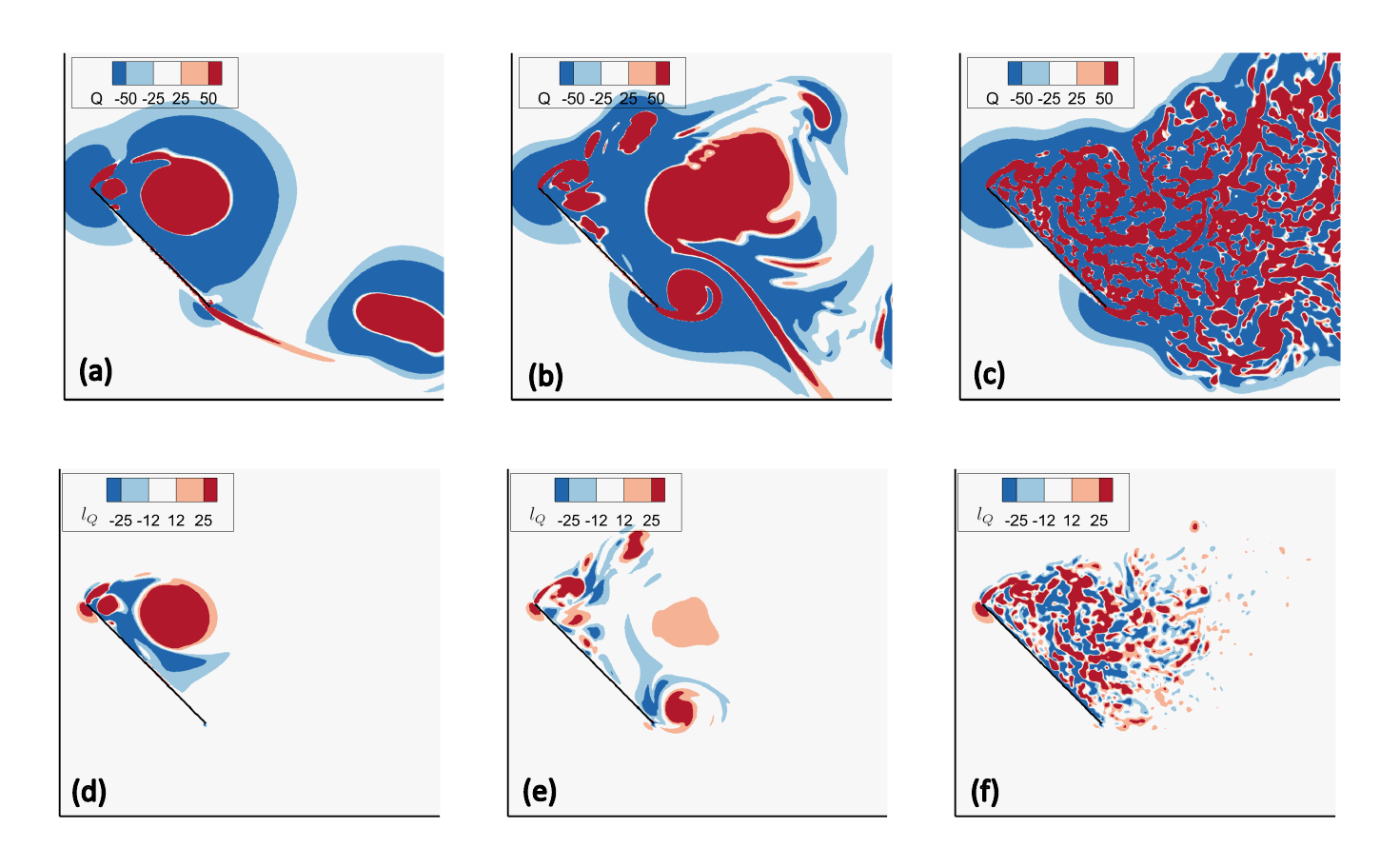}
\caption{{A spanwise averaged slice between $r/R=0.62 - 0.83$ is used to show the contour of $Q$ at three key time instance- (a) $t/T=0.048$, (b) $t/T=0.109$, and (c) $t/T=0.213$  . The corresponding vortex-induced lift force density ($l_Q$) is also shown at (d) $t/T=0.048$, (e) $t/T=0.109$, and (f) $t/T=0.213$.}}
\label{fig:thin_slice_lift}
\end{figure}

Figure \ref{fig:15slice_lift} shows the spanwise variation of sectional pressure lift across the span at the key time-instances that are our focus here. From this plot we note that a significant proportion of the total difference in lift at the key  time-instances is associated with the spanwise region  $0.62< r/R < 0.83$. Note that $t/T=0.048$ seems to be an outlier with the large sectional lift near the tip, but that increased sectional lift is compensated partially by the lower sectional lift for $r/R < 0.6$. Thus, we initially focus on this region of the blade and figure \ref{fig:thin_slice_lift}(a--c) shows a series of plots of the $Q$-field that is spanwise averaged over the region $r/R$ of 0.62 to 0.83. We note that at  $t/T$ = 0.048 where the first peak in the lift is found, the $Q$-field is dominated by a large LEV, a strong TEV and a small secondary vortex near the LEV that is induced by the reverse flow of the large LEV. As pointed by \cite{menon2021a}, there are significant regions of negative $Q$ that surround each vortex core and these regions induce a positive pressure force on the body, as opposed to the negative (suction) pressure force induced by the vortex cores with positive $Q$. In fact, \cite{menon2021a} noted that $\int_V{Q}dV$ is theoretically zero for flow around an immersed body. 
This integral might be non-zero but small in both simulations and experiments due to truncation and measurement errors respectively.
Notwithstanding this error, we expect an overall balance between positive $Q$ and negative $Q$ in the flow domain and the non-zero vortex-induced force emerges from the difference in the \emph{distribution} of these two quantities, and particularly, their proximity to the surface. Indeed, this proximity is exactly what is encoded in the influence field $\phi$ and multiplication of $Q$ with $\phi$ results in the vortex force-density which directly quantifies the force induced by every elemental region of the flow. Figure \ref{fig:thin_slice_lift}(d--f) shows contours of $l_Q$ corresponding to this spanwise segment averaged $Q$ field and we note that the multiplication with $\phi$ eliminates the TEV and also diminishes the outer corona of negative $Q$ around the LEV. This results in a region of positive $l_Q$ that is larger and more intense than that for negative $l_Q$. This observation can be connected with the presence of high lift at this time instance. 

However, the above correlation is still observational in nature and this observational correlation become slightly more difficult to extend to $t/T=0.109$ where the vortex field is significantly more complicated. At this time, the LEV is in the process of shedding from rotor blade and the negative $Q$ regions of both the LEV and the TEV combine to dominate the region around the blade. The $l_Q$ field for this time instance is highly fragmented making it difficult to draw clear conclusions, although the drop in lift coinciding with LEV shedding conforms to our understanding of these flows. Interestingly, \cite{otomo2021unsteady} noted the difficulty of predicting the lift precisely for this  stage in the vortex shedding process.

The situation for $t/T=0.213$ is much worse in this regard since the flow has already transitioned to turbulence and it is not possible to make observational deductions from the $Q$ or the $l_Q$ field.  Thus, while the FPM improves our ability to focus on the observable ($l_Q$) that determines the pressure lift, we need additional ways to quantify the strength and proximity of the pressure inducing structures. 

Here we introduce a simple data-driven method for separating and quantifying the effects of vortex strength from vortex proximity and extracting greater insights from the FPM as applied to these highly transient vortex dominated flows. We define $Q$-weighted influence fields for regions of positive and negative $Q$ as follows:
\begin{equation}
    \hat{\phi}_{\pm}= \frac{-2 \int_{V} \rho \phi \, Q_\pm \, dV}{\int_{V} \rho  Q_\pm dV} = \frac{L_{Q_\pm}}{\int_{V} \rho Q_
    \pm dV} \,\, ,
\end{equation}
where, $Q_\pm (\boldsymbol{x})=\frac{1}{2}\left(Q (\boldsymbol{x})\pm|Q(\boldsymbol{x})| \right)$ denotes the value of $Q$ in an elemental volume that contain either positive or a negative $Q$. The metrics $\hat{\phi}_{\pm}$ may be viewed as factors that are proportional to the averaged proximity of the positive and negative regions of the $Q$ field to the body of interest ($B$). Thus, higher (lower) values of these proximity function increase (decrease) the induced pressure forces from these two partitions of the $Q$ field. 
Now a metric for the overall strength of $Q$ in the domain called "$Q$-strength" can be defined by $\hat{Q} = \int_{V} \rho Q_+ dV= -\int_{V} \rho Q_- dV$, then by definition:
\begin{equation}
L_Q = L_{Q_+}+L_{Q_-} = \hat{Q}\left( \hat{\phi}_{+} - \hat{\phi}_{-} \right) = \hat{Q} \Delta \hat{\phi}_{\pm} 
\end{equation}
The above expression for vortex-induced lift therefore separates the effect of the net proximity from the body of the regions with positive and negative $Q$ (via the expression in the parenthesis) from the effect of the overall $Q$-strength ($\hat{Q}$), and the rise and fall of the lift force can now be examined with-respect-to the distinct variations of these two effects. We note that \cite{jardin2021empirical} showed that for low Reynolds number 2D flows, reasonable estimates of the lift on accelerating wings could be obtained using the circulation and position of the LEV, and this implicates $Q$-strength and location as the key parameters for lift as well.

\begin{figure}
\centering
\includegraphics[width=\textwidth]{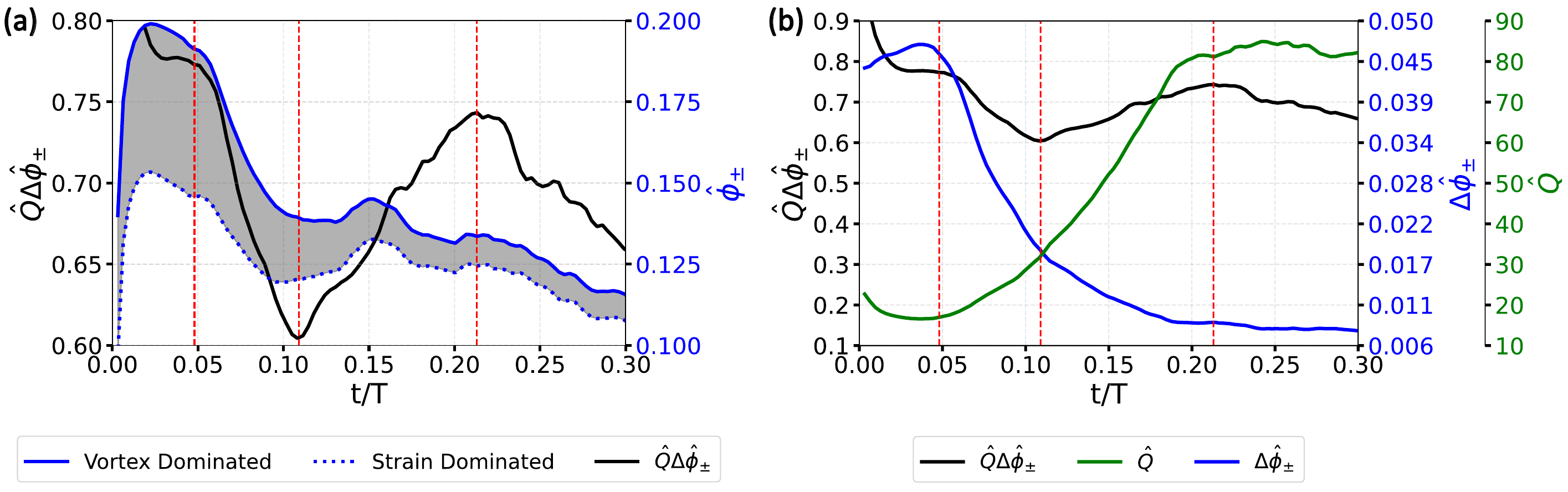}
\caption{(a) The weighted average of $\phi$ using $Q$ is denoted by $\hat{\phi}_{\pm}$ and is shown for strain ($\hat{\phi}_{-}$) and vortex dominated ($\hat{\phi}_{+}$)  region. The net force is shown by multiplying $\Delta\hat{\phi}_{\pm}$ with $\hat{Q}$. (b) The temporal variation of two components, $\hat{Q}$ and $\Delta \hat{\phi}_{\pm}$ and their product are shown. Both in this figure and in figure \ref{fig:Q_tavg}(a),  $\hat{\phi}_{\pm}$ is normalized by $c$, $\hat{Q}$ is normalized by $0.5\rho v_t^2 \zeta$ and $\hat{Q}\Delta\hat{\phi}_{\pm}$ is normalized by the force coefficient ($0.5\rho v_t^2 A_B$).}
\label{fig:phi_hat_temporal}
\end{figure}
Figure \ref{fig:phi_hat_temporal}(a) shows the time variation of the two proximity functions and the time-variation of the vortex-induced lift force $L_Q$. This plot shows that indeed the two proximity functions have different time-variations although $\hat{\phi}_{+}$ is always larger than $\hat{\phi}_{-}$ indicating (as expected) that the lift is always positive. 
Figure \ref{fig:phi_hat_temporal}(b) shows the time variations of $\Delta \hat{\phi}_{\pm}$ and $\hat{Q}$ and this plot is highly revealing. We note that $\hat{Q}$ rises rapidly in the very early stages up to $t/T =0.20$ due to the continuous flux of vorticity from the body in response to the establishment of a pressure gradient along the body. Beyond this, the overall $Q$-strength levels off. In contrast, the proximity difference is initially large indicating that regions of positive $Q$ are closer to the body than regions of negative $Q$, but this quantity decreases rapidly. The combination of these two curves explains the time-variation of the lift force. The first peak in lift at $t/T=0.048$ coincides with the peak in $\Delta \hat{\phi}_{\pm}$. Thus, while the overall vorticity strength is relatively low, the high proximity of the positive $Q$ (associated with the first LEV) at this time generates a large magnitude of lift.  Beyond this time, the proximity difference falls rapidly as the LEV starts to shed. However, the $Q$-strength is rising at this time and these competing effects result in the lift minimum at $t/T=0.109$. Beyond this, the lift rises due to the increasing $Q$-strength and around $t/T=0.2$, the growth in $\hat{Q}$ levels of and the continual slow decrease in the proximity difference starts to assert itself, thereby generating the peak at $t/T=0.213$. Thus the various extrema in this highly transient lift behavior for this complex 3D flow can be explained via the interplay between $Q$-strength and vortex proximity.  

\begin{figure}
\centering
     \includegraphics[width=1\textwidth]{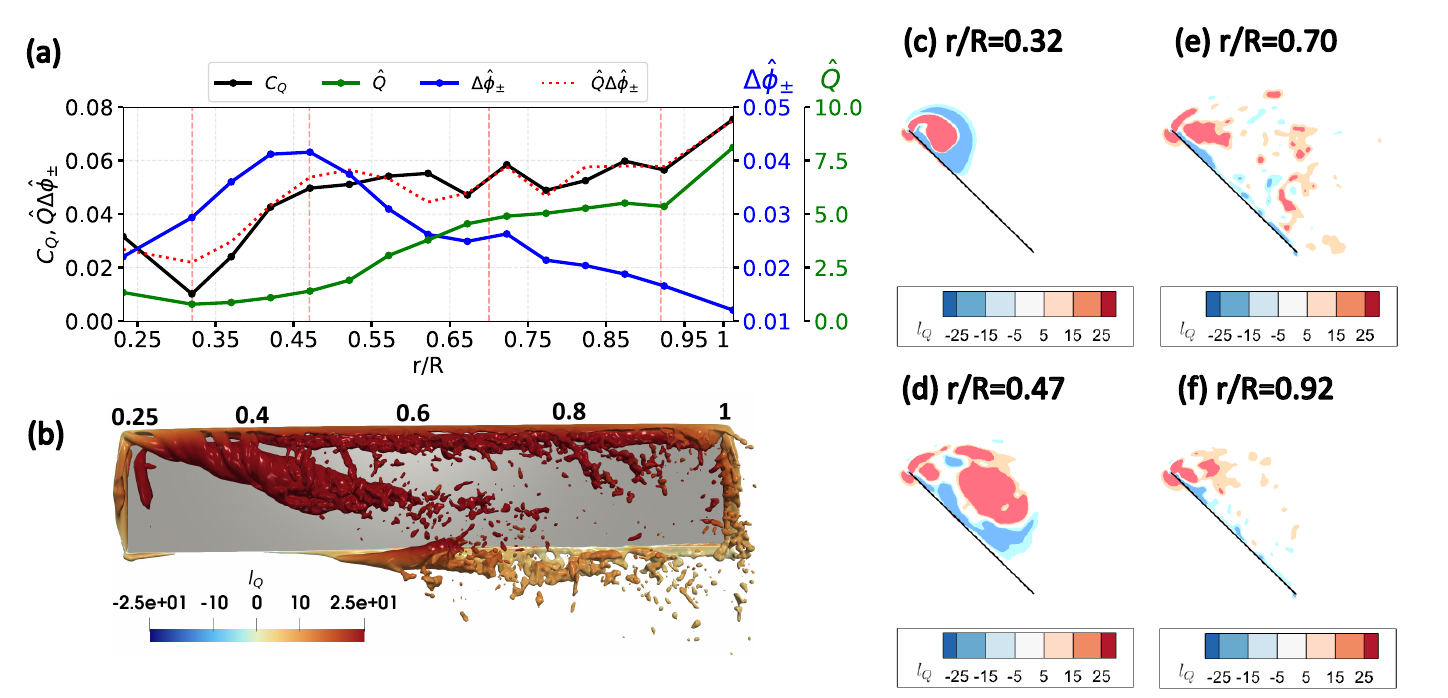}
\caption{(a) The spatial variation of $\hat{Q}$, $\Delta\hat{\phi}_\pm$, their product ($\hat{Q}\Delta\hat{\phi}_\pm$), and $C_Q$ is shown along the span of the rotor. The values are temporal averages between $t/T=$0 to 0.3. (b) Iso-surfaces of $Q$ temporally averaged between $t/T$= 0.0 to 0.3 and colored using $l_Q$. (c)-(f) 2D contours of $l_Q$ (also temporally averaged) are shown at four spanwise locations.}
\label{fig:Q_tavg}
\end{figure}

In the final section, we demonstrate that in addition to explaining the temporal variation in forces, this approach can also be used to examine the \emph{spatial} distribution of forces. The black line in figure \ref{fig:Q_tavg}(a) shows the spanwise variation of the sectional vortex-induced lift force on the blade, time-averaged between $t/T$=0 and 0.3, and we note that after an initial decrease, there is a rapid increase in the sectional lift up to $r/R$=0.36. Beyond this, the sectional lift is nearly constant till about $r/R=0.95$ and then increases in the last 5\% section of the rotor. The plot of $\hat{Q} \Delta \hat{\phi}_{\pm}$ follows the variation of the sectional vortex-induced lift very closely (it does not match exactly since we are not including the entire flow volume in the calculation of $\hat{Q} \Delta \hat{\phi}_{\pm}$) and thus, the spanwise variation in vortex-induced lift can indeed be understood by examining the $Q$-strength and the proximity functions. 

The sectional variation of $\hat{Q}$ and the proximity difference $\Delta \hat{\phi}_{\pm}$ are also plotted in the graph. The $Q$-strength drops slightly between $r/R$ of 0.25 and 0.32 (due to the tip vortex on internal tip, see figure \ref{fig:Q_tavg}(b)) and stays small up to $r/R$ = 0.54. It then rises more rapidly between $r/R$ of 0.54 and 0.7 where the primary LEV is growing and a secondary attached LEV as well as a trailing-edge vortex are also present, and then has a more gradual rise between $r/R$ of 0.7 and 0.92. Finally, the presence of the tip vortex results in a rapid increase in $\hat{Q}$ between $r/R$ of 0.92 and the tip. The contour of $l_Q$ is shown at four key spanwise locations in figures \ref{fig:Q_tavg}(c--f).

The proximity difference has a variation that is quite different from that of the $Q$-strength. This parameter is also initially small
but rises rapidly between $r/R$ of 0.25 and 0.5. This is connected with the presence of a very coherent and tightly rolled up attached LEV that is observed in many of the earlier vortex plots and in figure \ref{fig:Q_tavg}(b). Beyond  $r/R$ of 0.5, the proximity difference parameter falls continuously indicating that the LEV continues to move away from the rotor blade surface and regions of larger negative $Q$ form near the blade. Note from figure \ref{fig:Q_tavg}(b) that this is the spanwise location where the LEV first starts to exhibit a full breakdown. The variation of the sectional lift can now be explained based on these observation. The initial decrease of sectional lift between $r/R$ and 0.25 and 0.32 is due to drop in the $Q$-strength. Beyond this, the slight rise in $Q$-strength coupled with the rapid rise in the proximity difference lead to an increase in sectional lift until $r/R$=0.5, where this growth is stopped due to the reduction in the proximity difference. Between $r/R$ of 0.5 and 0.95, the opposite trends in the $Q$-strength (increasing) and proximity difference (decreasing) results in a plateau in the sectional lift. The final increase in sectional lift beyond $r/R$ of 0.95 is due to the increase in the $Q$-strength in the tip region.

\section{Conclusions}
Decades of research in fluid dynamics have revealed strong connections between the formation and evolution of vortical structures around immersed bodies and the unsteady forces they induce. These links often stem from the principle that vortex cores generate suction pressures, implicating features such as leading-edge vortices (LEVs), tip vortices, dynamic stall, flow separation, and vortex wakes. However, quite often, these connections are qualitative and leave significant room for ambiguity. In the current study, we have introduced a novel data-driven approach for establishing \emph{quantitative} and more definitive connections between vortical features and the unsteady forces induced on an immersed three dimensional body in high-Reynolds number flows. Using the force partitioning method, we derive two metrics -- the $Q$-strength and the vortex proximity parameter that enable us to provide quantitative measures of the effect of complex, dynamic, mutually interacting and evolving vortex structure on induced pressure forces. The method is demonstrated here by applying to simulation data for a revolving rotor blade at a relatively high Reynolds number, and explaining the temporal and spatial variations in lift.

While the current focus is on lift, the method readily extends to drag and aerodynamic moments \citep{menon2021a,mfpm_2025}. It is applicable to both 2D \emph{and} 3D flows, as vortex strength is quantified using the $Q$-field rather than circulation. In principle, the method can also be applied to experimental flow measurements, given prior successful implementation of FPM on PIV data \citep{zhu2023flow,zhu2023force}. This approach holds promise for informing design modifications or control strategies aimed at optimizing the performance of aerodynamic devices and control surfaces. 

Finally, and more broadly, the present work, in our view, reinforces the unique role of $Q$ in vortex identification. Unlike criteria such as vorticity or $\lambda_2$~\citep{jeong1995identification}, which are based purely on kinematic, $Q$ captures the \emph{action} of flow on bodies, devices and control surfaces immersed in flows, which provides a clear purpose for vortex identification.

\begin{bmhead}[Funding.]
Support from ARO (W911NF2120087) and ONR (N00014-24-12516) is acknowledged.
\end{bmhead}

\begin{bmhead}[Declaration of interests.]
The authors report no conflict of interest.
\end{bmhead}

\appendix
\section{Grid Convergence and FPM}
\label{app:gc}

\begin{figure}
         \centering
         \includegraphics[width=0.8\textwidth]{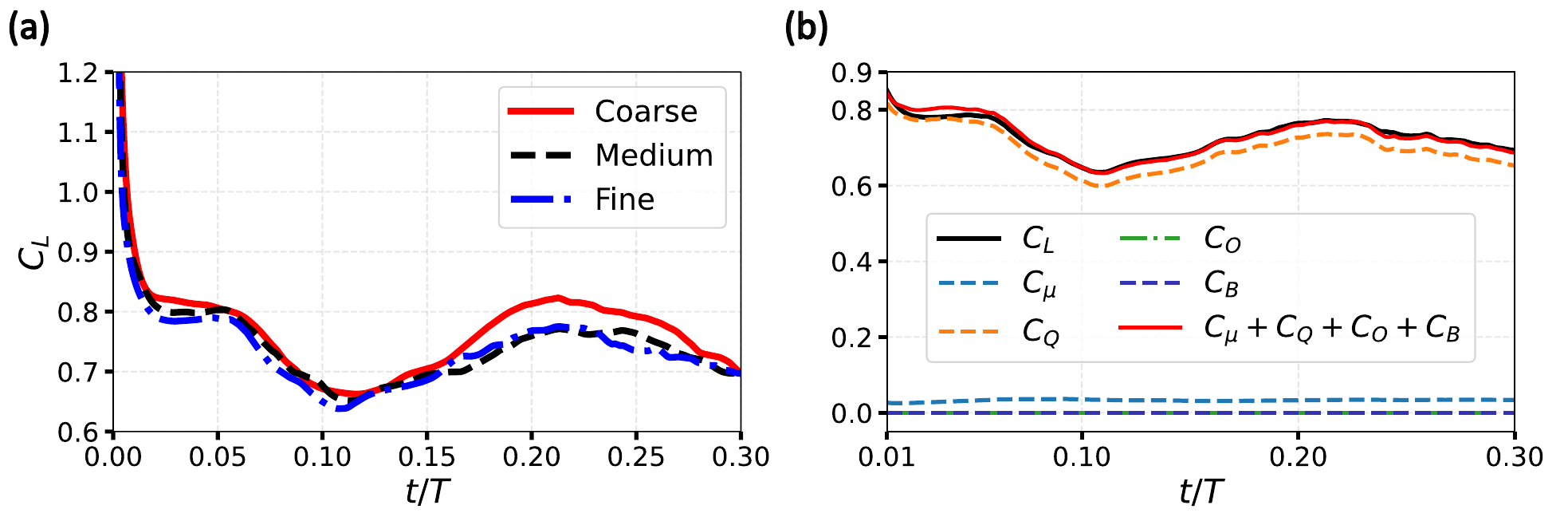}
\caption{(a) Grid convergence of lift coefficient shown for coarse, medium and fine mesh containing 8, 65 and 220 million grid points, respectively. (b) FPM application to fine mesh results - The total pressure lift ($C_L$) and the four force partitions; $C_\mu$: viscous diffusion, $C_Q$: vortex-induced, $C_O$: outer boundary, $C_B$: acceleration reaction (all normalized based on the tip velocity and blade area) versus time normalized by revolution period. 
}
\label{fig:grid-convergence}
\end{figure}

We employ a set of three meshes - coarse, medium and fine mesh, with 8, 65 and 220 million grid points respectively. The coefficient of the lift force is shown in figure \ref{fig:grid-convergence}(a) for 0.3 revolutions and we note that the results are well converged on fine mesh with the mean value of percent difference being around 3.9\% between the coarse and medium mesh and 0.31\% between the medium and fine mesh. The lift extrema, which are the focus of the current paper are also well converged on the fine mesh.

The fine mesh is used for all the results presented in this paper and this simulation consumed a total of approximately 350,000 CPU hours. The four force partitions (discussed in detail in \cite{menon2021a,mfpm_2025}), are shown in figure \ref{fig:grid-convergence}(b) and we note that about 95.5\% of the lift comes from the vortex-induced force while the viscous diffusion of the momentum term accounts for an average of 4.5\% of the net lift force. All other components are negligible (see \cite{menon2021c,mfpm_2025}).

\end{document}